\documentclass[notitlepage,prd,twocolumn,showpacs,preprintnumbers,nofootinbib,superscriptaddress,floatfix,tightenlines]{revtex4-1}
\usepackage{mathrsfs}
\usepackage{amssymb}
\usepackage{amsmath}
\usepackage{graphicx}
\usepackage{amsfonts,amsbsy}
\usepackage{nicefrac}
\usepackage{xcolor}
\usepackage[breaklinks,colorlinks,urlcolor=blue,citecolor=citcolor,linkcolor=lcolor]{hyperref}

\usepackage[utf8]{inputenc}

\definecolor{lcolor}{rgb}{0.5,0,0}
\definecolor{citcolor}{rgb}{0,0.3,0.0}
\usepackage{comment}

\newcommand{\mbf}{\mathbf}

\newcommand{\mrm}{\mathrm}
\newcommand{\ah}{\mrm{ah}}
\newcommand{\Tr}{\mrm{Tr}}

\newcommand{\HL}{\mrm{HL}}

\newcommand{\pInit}{p_0}
\newcommand{\Q}{Q}

\newcommand{\SU}{\mrm{SU}}
\newcommand{\su}{\mrm{su}}

\newcommand{\nha}{f}

\newcommand{\fE}{f_{\mrm{E}}}
\newcommand{\fpi}{f_{\pi}}

\newcommand{\fGen}{f}

\newcommand{\dpol}{d_{\mrm{pol}}}
\newcommand{\dA}{d_{\mrm{A}}}

\newcommand{\fig}{Fig.~}

\newcommand{\eq}{Eq.~}
\newcommand{\se}{Sec.~}
\newcommand{\eqs}{Eqs.~}
\newcommand{\re}{Ref.~}
\newcommand{\res}{Refs.~}
\newcommand{\app}{App.~}
\newcommand{\nr}[1]{(\ref{#1})}

\newcommand{\ud}{\mathrm{d}}

\newcommand{\pToFigs}{.}

\newcommand{\tDTwMi}{2D}
\newcommand{\tDThrMi}{2D+sc}
\newcommand{\tDThrIso}{3D}

\newcommand{\qs}{Q_{\mathrm{s}}}
\newcommand{\nc}{N_{\mathrm{c}}}

\newcommand{\lqcd}{\Lambda_{\mathrm{QCD}}}

\begin{document}

\title{Highly occupied gauge theories in $2+1$ dimensions: self-similar attractor}

\author{K.~Boguslavski} 
\affiliation{Institute for Theoretical Physics, Technische Universit\"{a}t Wien, 1040 Vienna, Austria}
\affiliation{Department of Physics, University of Jyv\"{a}skyl\"{a}, P.O.~Box 35, 40014 University of Jyv\"{a}skyl\"{a}, Finland}

\author{A.~Kurkela} 
\affiliation{Theoretical Physics Department, CERN, Geneva, Switzerland}
\affiliation{Faculty of Science and Technology, University of Stavanger, 4036 Stavanger, Norway}

\author{T.~Lappi} 
\affiliation{Department of Physics, University of Jyv\"{a}skyl\"{a}, P.O.~Box 35, 40014 University of Jyv\"{a}skyl\"{a}, Finland}
\affiliation{Helsinki Institute of Physics, P.O.~Box 64, 00014 University of Helsinki, Finland}

\author{J.~Peuron} 
\affiliation{European Centre for Theoretical Studies in Nuclear Physics and Related Areas(ECT*) and
Fondazione Bruno Kessler, Strada delle Tabarelle 286, I-38123 Villazzano (TN), Italy}

\preprint{CERN-TH-2019-115}

\begin{abstract}
 Motivated by the boost-invariant Glasma state in the initial stages in heavy-ion collisions,
 we perform classical-statistical simulations of SU(2) gauge theory in 2+1 dimensional 
 space-time both with and without a scalar field in the adjoint representation. 
 We show that irrespective of the details of the initial condition, the far-from-equilibrium evolution of these highly occupied systems approaches a unique universal attractor at high momenta that is the same for the gauge and scalar sectors. 
 We extract the scaling exponents and the form of the distribution function close to this non-thermal fixed point. We find that the dynamics are governed by an energy cascade to higher momenta with scaling exponents $\alpha = 3\beta$ and $\beta = -1/5$. 
 We argue that these values can be obtained from parametric estimates within kinetic theory indicating the dominance of small momentum transfer in the scattering processes.
 We also extract the Debye mass non-perturbatively from a longitudinally polarized correlator and observe an IR enhancement of the scalar correlation function for low momenta below the Debye mass. 
\end{abstract}

\maketitle


\section{Introduction}
\label{sec_introduction}

A characteristic feature of many highly-occupied systems is that they often approach universal self-similar attractors, also referred to as non-thermal fixed points (NTFP)~\cite{Micha:2004bv,Berges:2008wm}. Examples have been found with classical field methods in various theories in three spatial dimensions (\tDThrIso) including non-Abelian gauge theories, relativistic and non-relativistic scalar field theories \cite{Berges:2008mr,Berges:2012ev,Kurkela:2012hp,Schlichting:2012es,Lappi:2016ato,Berges:2013eia,Berges:2013fga,Micha:2004bv,Berges:2008wm,Nowak:2010tm,Nowak:2011sk,Berges:2014bba,Berges:2015ixa,Orioli:2015dxa,Schmied:2018upn,Mazeliauskas:2018yef}, and in two-dimensional scalar systems \cite{Gasenzer:2011by,Schole:2012kt,Karl:2016wko,Deng:2018xsk}. Non-thermal fixed points have recently also been found experimentally in ultra-cold atom experiments~\cite{Prufer:2018hto,Erne:2018gmz}. A kinetic theory description of the underlying theory is often a natural way to explain the existence and properties of such fixed points \cite{Micha:2004bv,Arnold:2002zm,Kurkela:2011ti,York:2014wja,Orioli:2015dxa,Walz:2017ffj,Chantesana:2018qsb}. 
These NTFPs appear because the interaction rate of the initial conditions is faster than that of the final equilibrium state. Therefore, the system loses memory of its initial conditions faster than it reaches thermal equilibrium and, hence, stays in a state that is not thermal yet but does not remember details of its initial conditions.

Much less is known about two-dimensional (\tDTwMi) gauge theories\footnote{See, however, Ref.~\cite{Arnold:2007tr} where the soft degrees of freedom are treated as \tDTwMi, but the hard ones as \tDThrIso.}. Differently from the three-dimensional case where an effective kinetic theory has been formulated to leading order accuracy \cite{Arnold:2002zm}, infrared (IR) effects play a stronger role in \tDTwMi\ due to the lower dimensionality. As we will discuss,  the hard (thermal) loop (HL) treatment used to regulate the Coulomb divergence of elastic scatterings in \tDThrIso\ is insufficient  in the two-dimensional case. Thus, it is a priori not obvious whether or to what extent quasi-particle descriptions are applicable and whether the system can exhibit self-similar behavior.

Apart from these theoretical questions, this uncertainty has also conceptual consequences for our understanding of the thermalization (hydrodynamization) process in ultra-relativistic heavy-ion collisions. In this context, non-linear interactions of gluons produced at central rapidities have been argued to lead to a transverse momentum scale $\qs \gg \lqcd$ up to which gluonic fields are of order $\mathcal{A} \sim 1/g$~\cite{Gelis:2010nm}, where $g$ is the gauge coupling. If this saturation scale is sufficiently hard, the system is weakly coupled $\alpha_s(Q_s) \equiv g^2/(4\pi) \ll 1$ and consists of highly occupied ``Glasma'' color fields ~\cite{Lappi:2006fp}. These are initially approximately boost-invariant along the beam axis and can be described by a \tDTwMi\ classical Yang-Mills field theory. 
Therefore, it is interesting to ask
to what extent hard-loop theory and quasi-particle approximations are applicable also to extremely anisotropic media and to \tDTwMi\ theories.
Adjacent to this is the question of what is the earliest time during the thermalization process when kinetic theory can be used to describe the dynamics.

In this paper, we will study whether highly occupied two-dimensional gauge theories approach a universal self-similar attractor and whether their scaling properties can be understood with simple kinetic theory arguments. 
We will consider two related SU(2) gauge theory systems using a classical lattice formulation.%
\footnote{We expect the results to carry over to SU(3) theories as well. The qualitative agreement of weakly-coupled SU($\nc$) theories far from equilibrium for $\nc = 2, 3$ has been observed for different phenomena \cite{Ipp:2010uy,Berges:2017igc,Berges:2008zt}.}
The first one of these systems is a 2+1-dimensional gauge theory (\tDTwMi). The second theory also includes a scalar field in the adjoint representation of the gauge group, and will be denoted as ``\tDThrMi''. The latter corresponds more closely to the situation in the initial effectively two-dimensional stage of a heavy-ion collision, since it is the theory one obtains by dimensional reduction starting from a 3+1-dimensional pure gauge theory and restricting it to field configurations (and gauge transformations) that do not depend on the longitudinal spatial coordinate. 

Our main result is that we indeed observe a self-similar scaling behavior for the hard modes of both theories 
that can be explained using parametric considerations in kinetic theory. Some evidence for such scaling behavior was seen in~\cite{Lappi:2017ckt}, where the focus was more on the determination of the plasmon frequency, and in the present work we establish with different methods the existence of the NTFP. While we focus here on the dynamics of hard modes, questions concerning HL and quasi-particle descriptions of soft momentum modes $p \sim m_D$ will be studied with unequal-time correlation functions in classical-statistical simulations in a forthcoming work.

This paper is structured as follows. In Sec.~\ref{sec_theory} we will briefly discuss the two theories we are studying, the initial conditions used and the numerical algorithm.  Then in Sec.~\ref{sec:results} we present the results from the numerical calculations. In Sec.~\ref{sec_Boltzmann_class} we derive the observed scaling exponents from a kinetic description. We conclude and outline some potential future work in Sec.~\ref{sec:conclu}. The Appendices cover details of our approach and of our analysis.


\section{Theoretical background}
\label{sec_theory}

\subsection{Theories and initial conditions}

We consider non-Abelian $\SU(\nc)$ gauge theories with $\nc = 2$ in $d=2$ spatial dimensions. The starting point is the classical gauge field action
\begin{align}
 \label{eq_class_action}
 S_{\mrm YM}[A] = -\frac{1}{4}\,\int d^{d+1}x\;F_a^{\mu\nu} F^a_{\mu\nu},
\end{align}
with $F^a_{\mu\nu} = \partial_\mu A^a_\nu - \partial_\nu A^a_\mu + g\, f^{abc}\, A^b_\mu A^c_\nu$, where repeated color indices $a = 1, ..., \nc^2 - 1$ and Lorentz indices $\mu, \nu = 0, ..., d$ imply summation over them. Using the generators $\Gamma^a$ of the $\su(\nc)$ algebra, the gauge field in fundamental representation reads $A_\mu = A^a_\mu \Gamma^a$.

We study the following two theories:
\begin{description}
 \item[\tDTwMi] gauge theory in $d = 2$ spatial dimensions, with the Yang-Mills action  \eqref{eq_class_action}.
 \item[\tDThrMi] the same gauge theory supplemented with an additional scalar field in the adjoint representation of the gauge group. This corresponds to a classical action
 \begin{align}
  S^{\tDThrMi}_{\mrm YM}[A] = S_{\mrm YM}^{2D}[A] + S_{\phi}^{2D}[\phi]
 \end{align}
with an adjoint scalar field $\phi^a$ and 
\begin{align}
 S_{\phi}^{2D}[\phi] = -\frac{1}{2}\,\int d^{2+1}x\;(D^{ab}_{j}\phi^b) (D_{ac}^{j}\phi^c).
\end{align}
Here the summation is over $j = 1,2$ and the covariant derivative is $D^{ab}_j = \delta^{ab}\partial_j - g f^{abc} A_j^c$. This theory can be obtained from Yang-Mills theory in 3 spatial dimensions by dimensional reduction, assuming that the field configurations do not depend on the  third coordinate $x^3$. To maintain this symmetry, also gauge transformations are not allowed to depend on $x^3$, turning the third component of the gauge field into a scalar 
$A_3^a \equiv \phi^a$.
\end{description}
Note that in 2D the dimensionalities of the fields  and coupling constants are different from the 3D case. The action must be dimensionless $[S_{\mrm YM}] = [S_{\mrm YM}^{2D}] = [S_{\phi}^{2D}] = 0$, from which one easily deduces that  $[g] = 1/2$ and $[A] = [\phi] = 1/2$. The dimensionality of the interaction term of the covariant derivative has to be that of a derivative: $[gA] = 1$, as in three spatial dimensions. 

The theory \tDThrMi\ is the nonexpanding space-time analogy of the boost-invariant Glasma, while the \tDTwMi\ theory also drops the adjoint scalar contribution. Therefore, both theories can be considered as simplifications of the Glasma state. Note that there is only $\dpol = 1$ transverse polarization in \tDTwMi, while the \tDThrMi\ theory, originating from a \tDThrIso\ system, has $\dpol = 2$ transverse polarizations: one from gauge degrees of freedom and one from adjoint scalars.

The systems are initialized, using the method described in \se\ref{sec_theory_classical}, with a field configuration that has a chosen single-particle distribution function $f(t,\mbf p)$ at the initial time $\Q t = 0$. 
Here $\Q$ is a conserved  momentum scale characterizing the system and will be defined in \eqref{eq_Q_def}.
We consider weakly coupled $g^2/\Q \ll 1$ but highly 
occupied $f \gg 1$ initial conditions of the form 
\begin{align}
 \label{eq_2D_IC}
 f(t=0, p) = \frac{\Q}{g^2}\, n_0\, e^{-\frac{p^2}{2\pInit^2}},
\end{align}
for gauge and scalar fields.
Unless stated otherwise we will use $n_0 = 0.1,$ for which $\pInit = \Q$ for our chosen definition as detailed below. As we will show in Sec.~\ref{sec:results}, the exact form of the initial conditions and the values of $\pInit$ and $n_0$ separately are not relevant after a transient time, since the systems will approach an attractor solution that only depends on $\Q$. 
 
To define the characteristic momentum scale $\Q$, note that the energy density is a conserved quantity in the systems studied here 
and can be computed in a gauge-invariant way in classical-statistical field theory (e.g., in \cite{Kurkela:2012hp,Berges:2013fga}).
The combination that  we have access to in the classical field formulation is  the energy density scaled with the coupling $g^2 \varepsilon$, which has the momentum dimension $[g^2\varepsilon] = 4$. 
This allows us to define a constant momentum scale in a gauge-invariant way as 
\begin{align}
 \label{eq_Q_def}
 \Q \equiv \sqrt[4]{\frac{C\,g^2 \varepsilon}{\dpol\, \dA}}\,,
\end{align}
with the number of transverse polarizations $\dpol$, and the dimension $\dA = \nc^2-1$ of the adjoint representation. The constant $C$ is taken as $C = 20 \sqrt{2\pi} \approx 50$, a value that has no physical meaning and has been chosen for convenience such that for $n_0 = 0.1$ one indeed has $\pInit = \Q$. 
Recall that the coupling constant $g$ is now dimensionful: if one keeps the \emph{dimensionless} combination $g^2/Q$ constant, it is easy to see that \nr{eq_Q_def} leads to the proportionality $\Q \propto \sqrt[3]{\varepsilon}$ that is natural for a scale derived from a 2-dimensional energy density.
The scale $\Q$ will be used to measure all dimensionful quantities.

\subsection{Semi-classical simulations}
\label{sec_theory_classical}

At high occupation numbers, we can use the classical-statistical approximation to study the dynamical evolution of systems far from equilibrium \cite{Aarts:2001yn,Smit:2002yg}. 
A description of this standard technique can be found, for instance, in \res\cite{Berges:2013fga,Boguslavski:2018beu}. 
Then all fields are classical and discretized on a cubic lattice of size $N_s^2$ with lattice spacing $a_s$. The real-time dynamics results from solving a gauge-covariant formulation of the classical Hamiltonian equations of motion in temporal gauge $A_0 = 0$ in a leapfrog scheme for the gauge-covariant link fields $U_j(t,\mbf x) \approx \exp(i a_s gA_j(t,\mbf x))$ and chromo-electric fields $E^j_a = \partial_t A_j^a$. For the theory \tDThrMi, we use $j = 1,2,3$ in the same scheme. 

The fields can be initialized in momentum space by requiring that the transversely polarized fields\footnote{The fields at each momentum $\mbf p = (p_1,p_2)$ can be split into a transverse and longitudinal part $A_j^a(\mbf p) = A_{T,j}^a(\mbf p) + A_{L,j}^a(\mbf p)$, such that $A_{T}^a(\mbf p) \cdot \mbf p = 0$ while $|A_{L}^a(\mbf p) \cdot \mbf p | = |A_{L}^a(\mbf p)|\, p$.}
at $\Q t = 0$ follow 
\begin{eqnarray}
 \dfrac{1}{\dA V} \langle A_T^a(0,\mbf p) (A_T^a(0,\mbf p))^* \rangle\, &=& \dfrac{f(0,p)}{p} 
 \label{eq_IC_fields1} \\
 \dfrac{1}{\dA V} \langle E_T^a(0,\mbf p) (E_T^a(0,\mbf p))^* \rangle\, &= &p\,f(0,p)  
 \label{eq_IC_fields2} \\
 \dfrac{1}{\dA V} \langle \phi^a(0,\mbf p) (\phi^a(0,\mbf p))^* \rangle\, &=& \dfrac{f(0,p)}{p} 
  \label{eq_IC_fields3} \\
 \dfrac{1}{\dA V} \langle \pi^a(0,\mbf p) (\pi^a(0,\mbf p))^* \rangle\, &= &p\,f(0,p) \,,
  \label{eq_IC_fields4} 
\end{eqnarray}
with $\pi^a \equiv E^3_a$, while other combinations for two-point functions vanish, as well as $\langle A \rangle = \langle E \rangle = 0$.%
\footnote{In practice, this is achieved by setting $A_j^a(t=0,\mbf p) = \sqrt{f(t=0, p)/p}\; c_a(\mbf p)\, v_{T,j}(\mbf p)$, and similarly for the other fields, with complex-valued Gaussian random numbers $c_a(\mbf p)$ and the transverse polarization vector $v_{T}(\mbf p)$.}

Since such initial conditions violate the Gauss law constraint, the latter is restored with the algorithm from \cite{Moore:1996qs} before starting the dynamical evolution. Alternatively, we also started with initial conditions with $E^j = 0$ but twice the amplitude $n_0$, where the Gauss constraint is automatically satisfied and the energy density approximately the same. We found that both lead to the same dynamics after a short transient time. 

We will be especially interested in observables in momentum space. For that, we fix the gauge to Coulomb-like gauge $\partial_j A_j = 0$ at the measurement time (see also \app\ref{app_IC_in_CoulombGauge}) and Fourier transform the fields according to $A(t,\mbf p) = \int d^dx\, e^{-i\mbf p \cdot \mbf x} A(t,\mbf x)$. A central quantity of interest is the single-particle distribution function $f(t,\mbf p)$, which provides the occupation number density of the system at a given time and momentum. One can define the distribution function using different correlators, such as in \eqref{eq_IC_fields1} -- \eqref{eq_IC_fields4}. Unless stated otherwise, our standard definition will be
\begin{align}
\label{eq_defin_fp}
 \fE(t,p) := \dfrac{\langle E_T E_T^* \rangle(t,p)}{\omega(p)} \\
 \fpi(t,p) := \dfrac{\langle \pi \pi^* \rangle(t,p)}{\omega(p)}\,,
\end{align}
where we will set the dispersion $\omega(p) = p$ neglecting soft scale effects since we are primarily interested in the dynamics at high momenta. We also used an abbreviated notation
\begin{align}
 \langle E_T E_T^* \rangle(t,\mbf p) = \dfrac{1}{\dA V} \langle E_T^a(t,\mbf p) (E_T^a(t,\mbf p))^* \rangle\,,
\end{align}
and similarly for other correlators. 

We note that the classical-statistical framework with the initial correlators \eqref{eq_IC_fields1} -- \eqref{eq_IC_fields4} corresponds to an accurate mapping of the corresponding quantum field theory onto a classical-statistical field theory in the limit of weak coupling $g^2 \rightarrow 0$ for $g^2 f(t,p)$ held fixed \cite{Aarts:2001yn,Smit:2002yg,Berges:2013lsa}. 
In this limit, the vacuum $1/2$ contribution to the distribution function is suppressed by $g^2$ and is negligible for the observables studied here. Therefore, contributions at the order of the lattice cutoff $p \sim 1/a_s \gg \Lambda$ are non-physical in our case and we use different discretizations to verify that our results are not sensitive to $a_s$ (or to the lattice volume).


\begin{figure}[t]
	\centering
	\includegraphics[scale=0.7]{\pToFigs/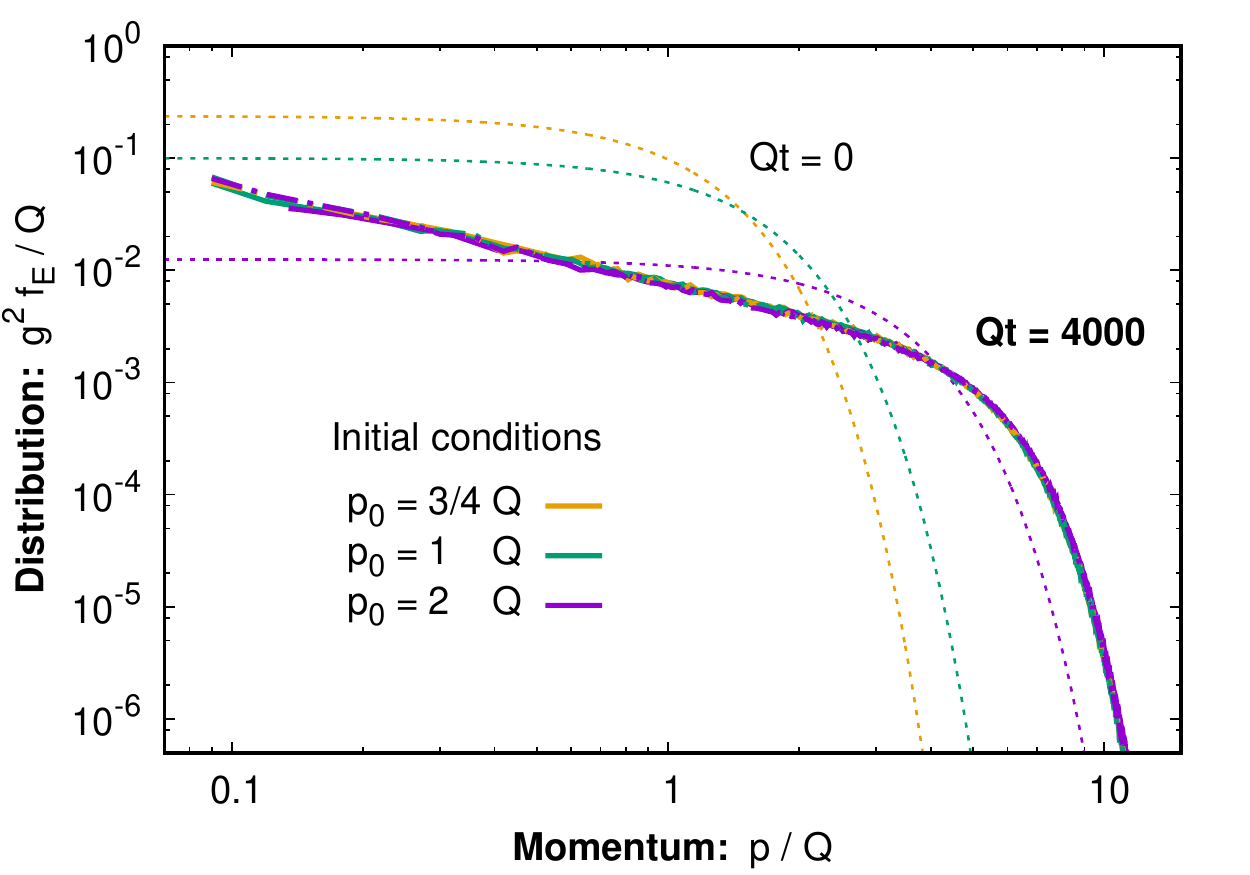}
	\caption{
	The gauge distributions $\fE$ at the initial time $\Q t = 0$  for three different initial conditions are shown by dashed lines; note that these same initial conditions are used for both \tDTwMi\ and \tDThrMi\ theories. At a later time $\Q t = 4000$, full lines show the distributions from these initial conditions in the  \tDTwMi\ theory and dashed-dotted lines in the \tDThrMi\ theory. These six curves overlap so well that they are indistinguishable in this plot, demonstrating the attractor property of the common non-thermal fixed point for both theories.	
	}
	\label{fig_fp_diffIC_2D}
\end{figure}

\section{Simulation results}
\label{sec:results}

\subsection{Universality and self-similarity}
\label{sec:results-selfsim}

In this section, we demonstrate numerically that starting from initial conditions with high occupation numbers, both theories \tDTwMi\ and \tDThrMi\ approach a common non-thermal fixed point at high momenta where the distribution function follows a self-similar evolution
\begin{align}
 \label{eq_selfsim}
 \fGen(t,p) = (\Q t)^{\alpha} f_s\left((\Q t)^{\beta} p \right) .
\end{align}
In order to constitute a universal non-thermal fixed point, the scaling exponents $\alpha$, $\beta$ and the scaling function $f_s(p)$ in \eq\eqref{eq_selfsim} must be  the same for different initial conditions. The time evolution at the fixed point only depends on a single conserved quantity, which is the energy density $\varepsilon$ in the case of an energy cascade to higher momenta \cite{Micha:2004bv,Berges:2013lsa,York:2014wja}, as is the case here. Hence, the distribution function becomes insensitive to details of the initial conditions after a transient evolution when rescaled with the only dimensionful scale $\Q$ determined by $\varepsilon$. 

This attractor property is illustrated in \fig\ref{fig_fp_diffIC_2D} where we show the gauge distribution $\fE$ as a function of momentum in a double-logarithmic plot for both theories. Dashed lines correspond to different initial conditions at $\Q t = 0$, where the fields are constructed to reproduce the chosen momentum distribution according to \eqs\eqref{eq_IC_fields1}--\nr{eq_IC_fields4}. Each initial condition was used in both theories and the figure shows their $\fE$  also at the later time $\Q t = 4000$, where dashed-dotted lines correspond to the \tDThrMi\ theory and full lines to the \tDTwMi\ theory. Although resulting from different initial conditions and theories, all six distributions at $\Q t = 4000$ lie indistinguishably on a single curve. This demonstrates that after some transient time that depends on details of the initial conditions, systems in both theories get close to the same attractor. There they follow a universal evolution, insensitive to their original initial conditions.

\begin{figure}[t]
	\centering
	\includegraphics[scale=0.7]{\pToFigs/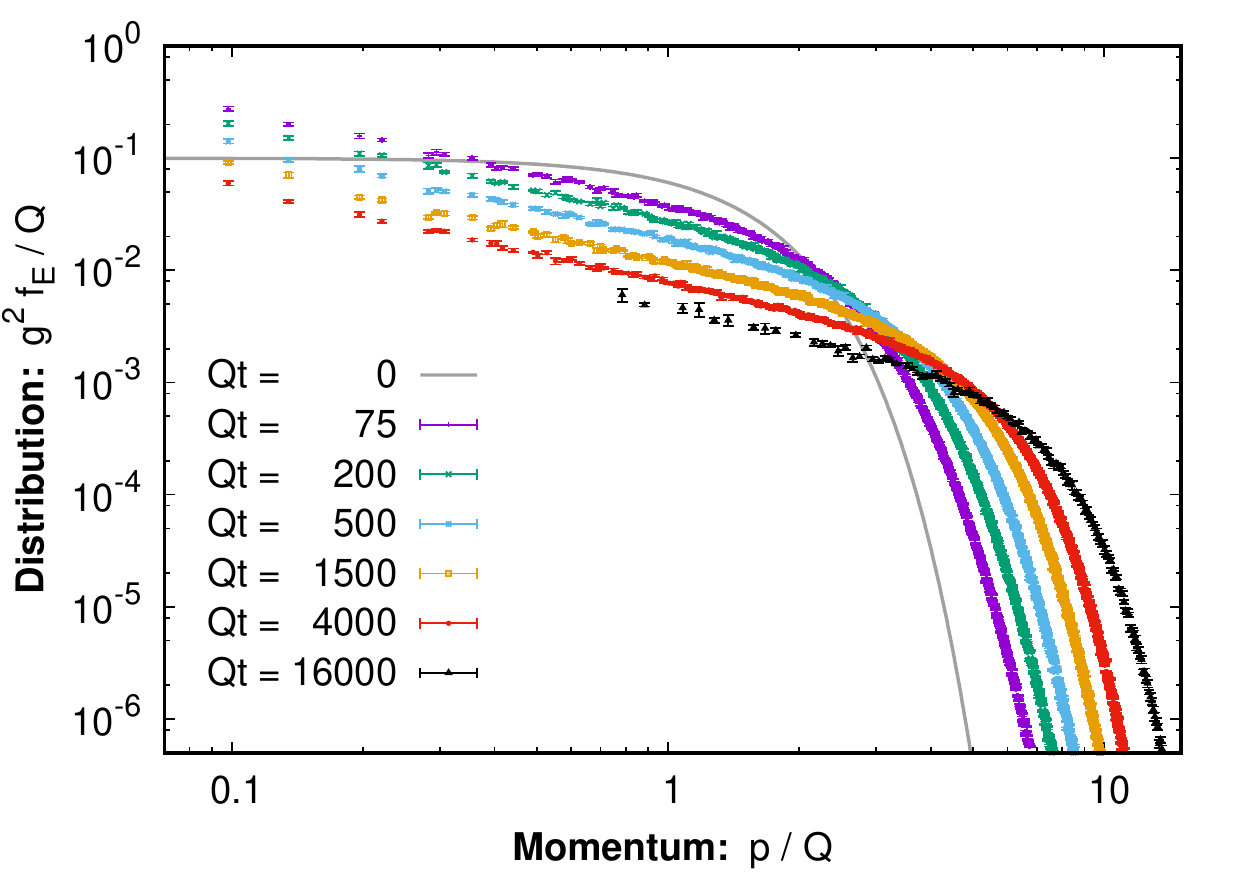}
	\includegraphics[scale=0.7]{\pToFigs/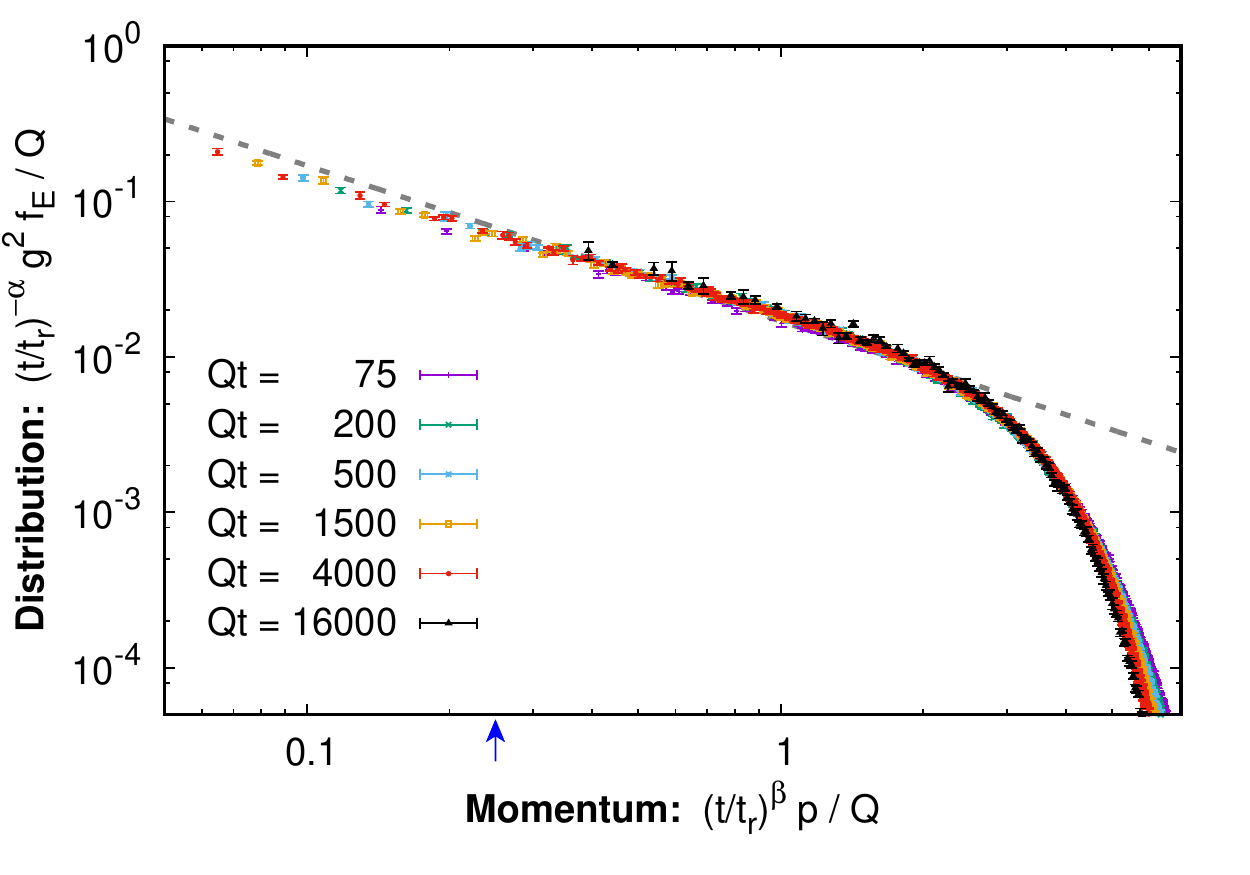}
	\caption{ Spectra prior to rescaling {\bf (top)} and rescaled occupation numbers and momenta {\bf (bottom)} are shown for the \tDTwMi\ theory at different times. For the rescaling, we used the reference time $\Q t_r = 500$ and the scaling exponents $\beta = -1/5$ and $\alpha = 3\beta$.
	The gray dashed line corresponds to a power law $p^{-1}$. The blue arrow indicates the maximal value of the Debye mass $m_D$ for the times displayed.}
	\label{fig_selfsim_2D}
\end{figure}

\begin{figure}[t]
	\centering
	\includegraphics[scale=0.7]{\pToFigs/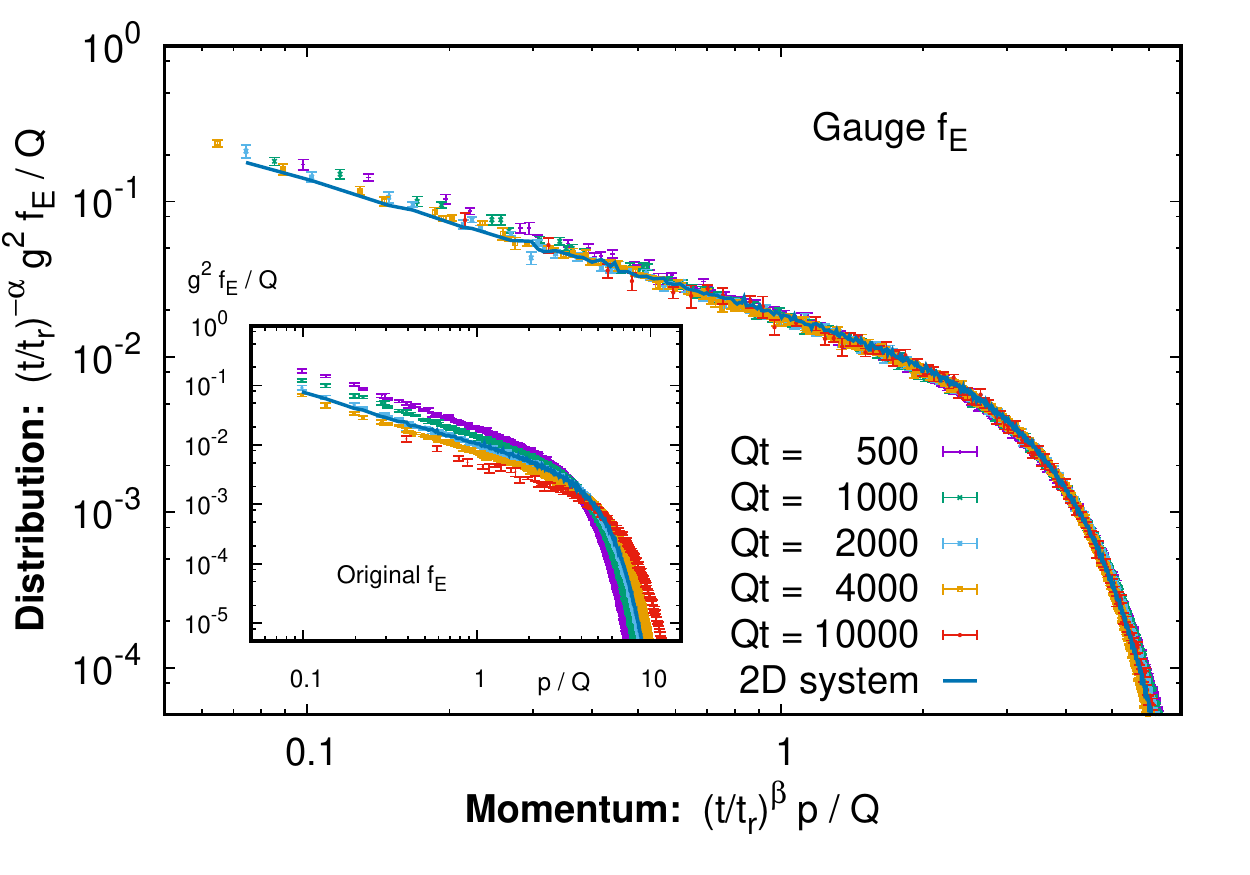}
	\includegraphics[scale=0.7]{\pToFigs/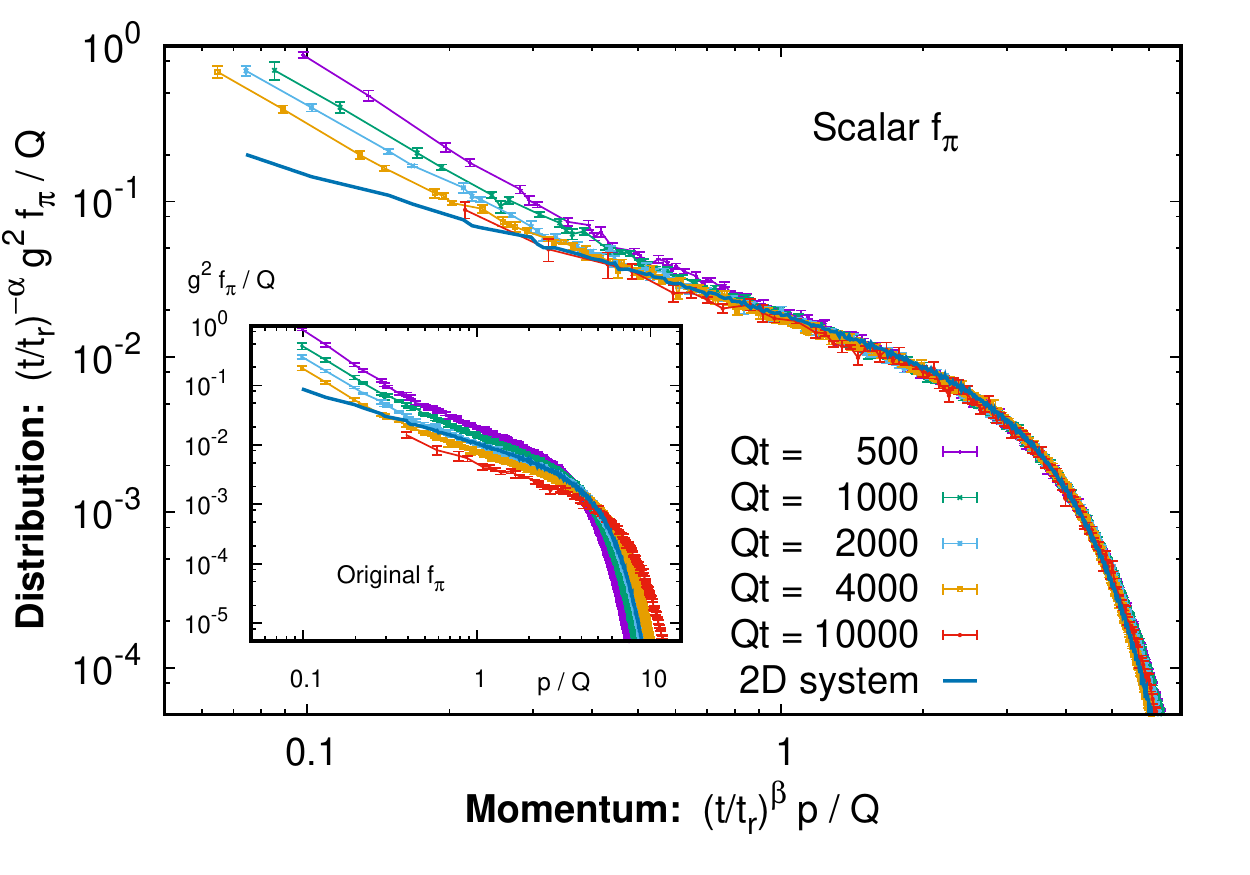}
	\caption{Self-similar evolution of the \tDThrMi\ theory for different times with rescaled occupation numbers and momenta for the gauge distribution $\fE$ {\bf (top)} and the scalar distribution $\fpi$ {\bf (bottom)}. The same values for $\alpha$ and $\beta$ as in \fig\ref{fig_selfsim_2D} are used. For comparison, we show the corresponding $\fE$ of the \tDTwMi\ theory for $\Q t = 2000$ as a dark-blue line.}
	\label{fig_selfsim_eff2D}
\end{figure}

In this universal regime, the distribution function becomes self-similar, following \eq\eqref{eq_selfsim}. This is demonstrated in \fig\ref{fig_selfsim_2D} for the \tDTwMi\ theory. The upper panel depicts the distribution function in the universal regime at several vastly different times%
\footnote{For $\Q t \leq 4000$ a $768^2$ lattice with lattice spacing $\Q a_s = 1/12$ has been used, for the later time we used a $256^2$ lattice with $\Q a_s = 1/16$, where the first two points of the latter were omitted due to volume artefacts. These artefacts occur when the lattice is too small to contain the screening mass $m_D$. In this situation, the smallest momentum modes are artificially enhanced. We checked that simulations of both discretizations coincide otherwise for $\Q t \leq 4000$.} 
$Qt=75-16000$. The lower panel shows this same
data in rescaled coordinates: the rescaled gauge distribution $(t/t_r)^{-\alpha} g^2 \fE / \Q$ is shown as a function of rescaled momentum $(t/t_r)^{\beta} p / \Q$. The scaling indices $\alpha$ and $\beta$ have been numerically extracted to produce the best overlap of the distribution functions at the different times employing a least-square fit procedure \cite{Berges:2013fga,Orioli:2015dxa} as detailed in \app\ref{app:likelihood}, leading to best-fit values
\begin{align}
\label{eq_scaling_data}
    \alpha_{\mrm{fit}} - 3\beta_{\mrm{fit}} \,&= ~~ 0.01 \pm 0.02 \\
\label{eq_scaling_data2}
    \beta_{\mrm{fit}} \,&= -0.19 \pm 0.015 \,.
\end{align}
The first combination results from energy conservation and that its best-fit value is consistent with zero is a consistency check of the procedure. For \fig\ref{fig_selfsim_2D} (as well as for all the following figures) we use the values that will be derived in Sec.~\ref{sec_Boltzmann_class}:
\begin{align}
\label{eq_scaling_values_plots}
    \alpha = 3\beta\,, \quad \beta = - \dfrac{1}{5} \,,
\end{align}
which are consistent with the extracted ones in \eqs\eqref{eq_scaling_data} and \eqref{eq_scaling_data2}. 
The good overlap of the different curves obtained at different times demonstrates scaling behaviour.

A similar conclusion can be drawn for the \tDThrMi\ theory. In \fig\ref{fig_selfsim_eff2D}, the rescaled gauge and scalar distributions $\fE$ and $\fpi$ are shown in the upper and lower panels, respectively.\footnote{For $\Q t \leq 4000$ a $512^2$ lattice with spacing $\Q a_s = 1/8$ has been used, for the later time we used a $256^2$ lattice with $\Q a_s = 1/16$. We checked that both discretizations coincide for $\Q t \leq 4000$.}
For comparison, the original curves are depicted in the insets. For the rescaling of amplitudes and momenta, the same exponents $\alpha$ and $\beta$ have been used as in \fig\ref{fig_selfsim_2D} for the \tDTwMi\ theory. One indeed observes that at high momenta the rescaled gauge distributions as well as the scalar curves lie on top of each other within error bars. This form also agrees with the \tDTwMi\ theory, which can be seen by comparison to the dark-blue curve.

The scaling function $f_s(p)$ of the gauge distribution consists of a power law $\propto (p/\Q)^{-\sigma}$ at lower momenta and a steep drop at high momenta. This closely resembles the non-thermal fixed point in \tDThrIso\ theory \cite{Berges:2008mr,Kurkela:2012hp,Berges:2013eia}, which also consists of a power law at low momenta and a steep decrease at high momenta.
The power law at small momenta is consistent with $\sigma=1$, which can be seen in the lower panel of \fig\ref{fig_selfsim_2D}, where a power law with $\sigma = 1$ is also displayed. 
Small deviations from this power law occur at very low momenta below the Debye mass that is indicated by the blue arrow.
This value $\sigma=1$, corresponds to a classical thermal distribution $T_{\mrm{eff}}/p$ at low momenta with a time-dependent effective temperature $T_{\mrm{eff}}$ \cite{Kurkela:2011ti}. Analytical considerations in effective kinetic theory suggest that the form of the distribution function in the infrared should take this form also out of equilibrium \cite{York:2014wja}.
However, numerical classical Yang-Mills simulations in \tDThrIso\ theory have found large corrections to this quantity at early times leading to extractions of $\sigma \approx 1.3$ \cite{Berges:2008mr,Kurkela:2012hp}. Such corrections to the classical thermal distribution seem to be absent in these 2D theories.

\begin{figure}[t]
	\centering
	\includegraphics[scale=0.7]{\pToFigs/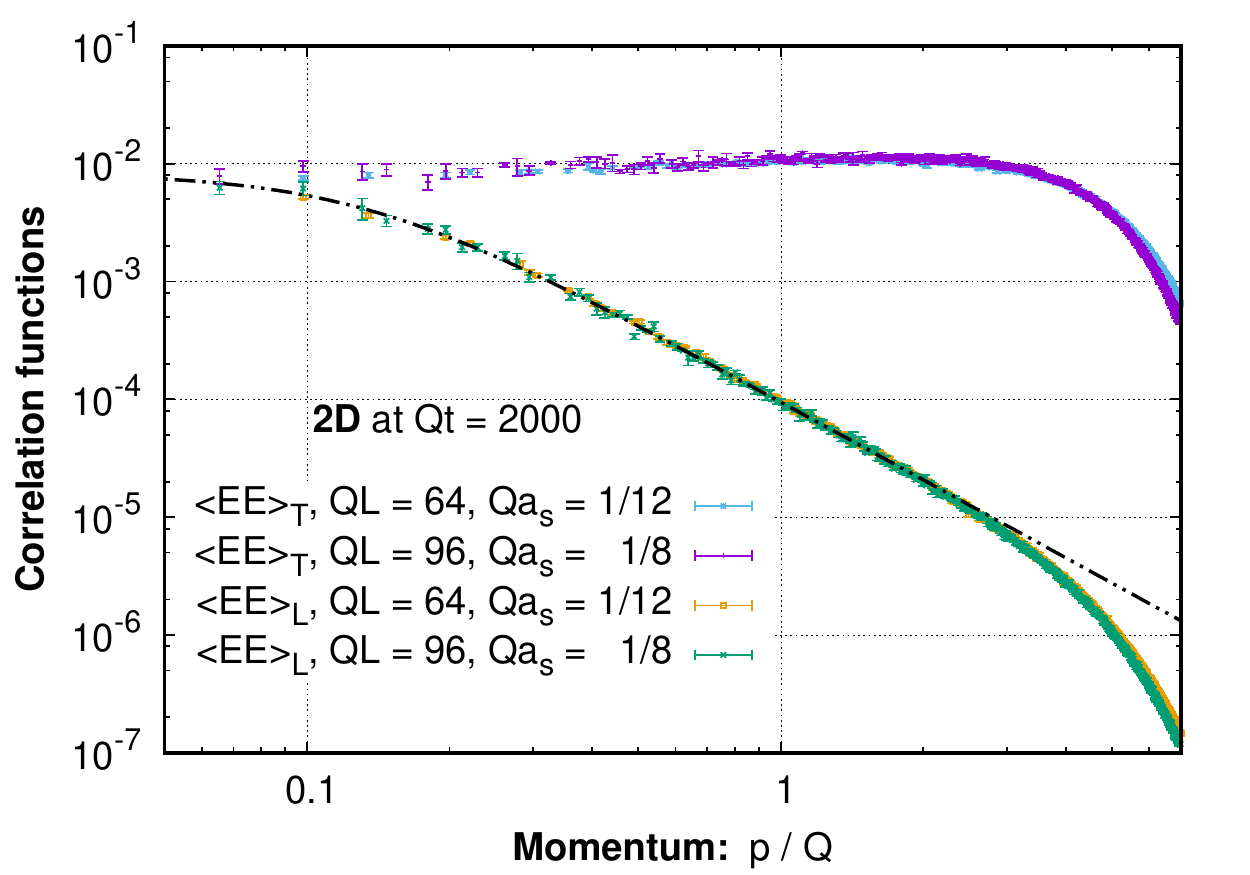}
	\includegraphics[scale=0.7]{\pToFigs/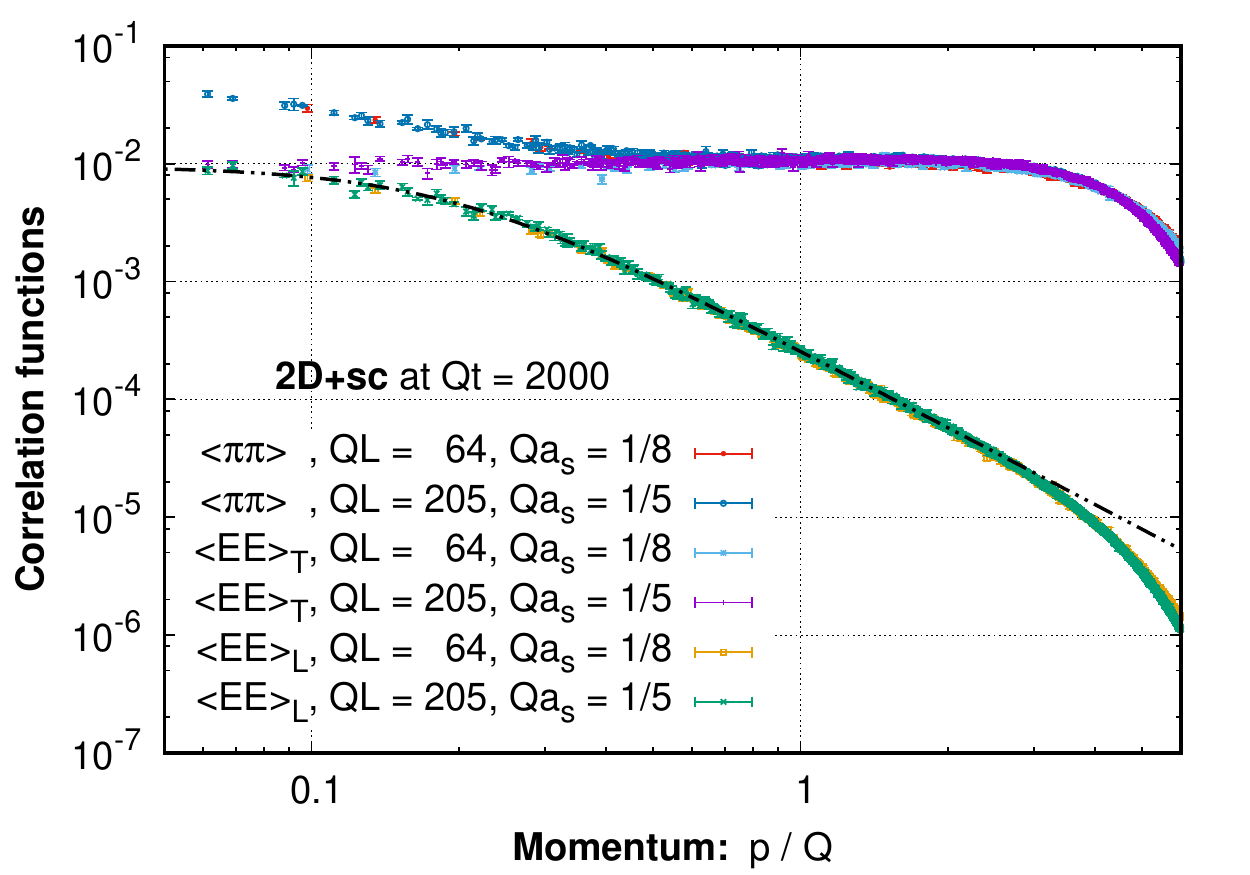}
	\caption{Correlation functions at $\Q t = 2000$ for \tDTwMi\ {\bf (top)} and \tDThrMi\ theories {\bf (bottom)} shown for different discretization parameters. Black dashed-dotted lines correspond to fitting $\langle E_L E_L^* \rangle$ to the functional form \eqref{eq_fit_long}.}
	\label{fig_TvsL}
\end{figure}

The main difference between gauge and scalar distributions is that the scalar curves $\fpi$ are enhanced for low momenta $p \lesssim m_D$ roughly below the Debye mass that will be discussed in \se\ref{sec:results-corr_mass}. This IR enhancement can be seen in the lower panels of \fig\ref{fig_selfsim_eff2D} and, further below, of \fig\ref{fig_TvsL}. It may seem similar to the case of $\mathcal{O}(N)$-symmetric scalar field theories where an IR region has been observed \cite{Berges:2008wm,Gasenzer:2011by} or may even be connected to 
non-trivial topological structures \cite{Gasenzer:2013era}. However, since this enhancement is not part of the self-similar region at high momenta, a detailed study is beyond the scope of this work and is left for further study  elsewhere.

\begin{figure}[t]
	\centering
	\includegraphics[scale=0.7]{\pToFigs/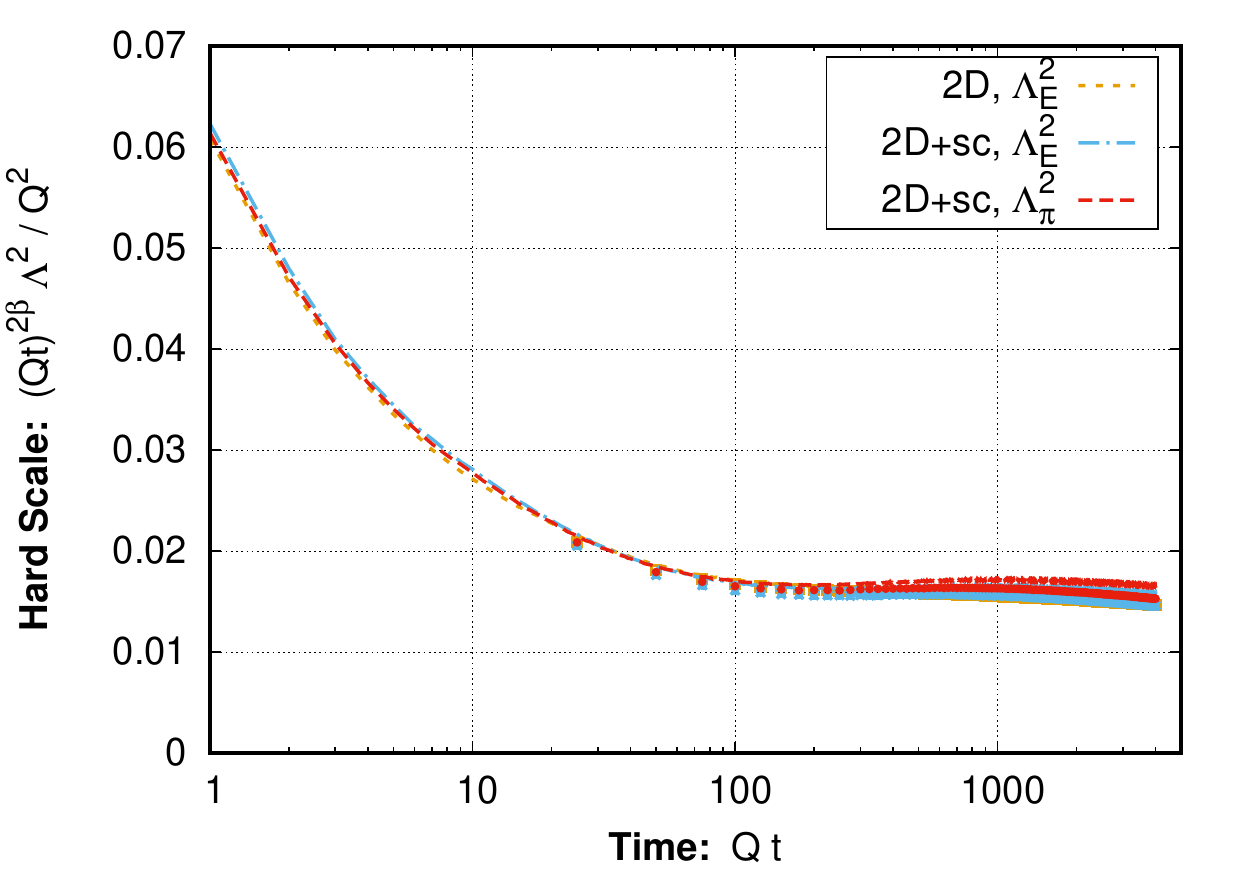}
	\caption{Different hard scales $\Lambda^2(t)$ for the \tDTwMi\ and \tDThrMi\ theories. 
The gauge-invariant definitions \eqref{eq_hardscale} and \eqref{eq_hardscale_pi} are shown with dashed lines compared to the perturbative integral expressions $\Lambda^2_{\mrm{pert}}(t)$ in \eqref{eq_hardscale_pert} as points. The curves are rescaled with $t^{2\beta}$ with $\beta = -1/5$.}
\label{fig_hardscale}
\end{figure}
	
\begin{figure}[t]
	\centering
		\includegraphics[scale=0.7]{\pToFigs/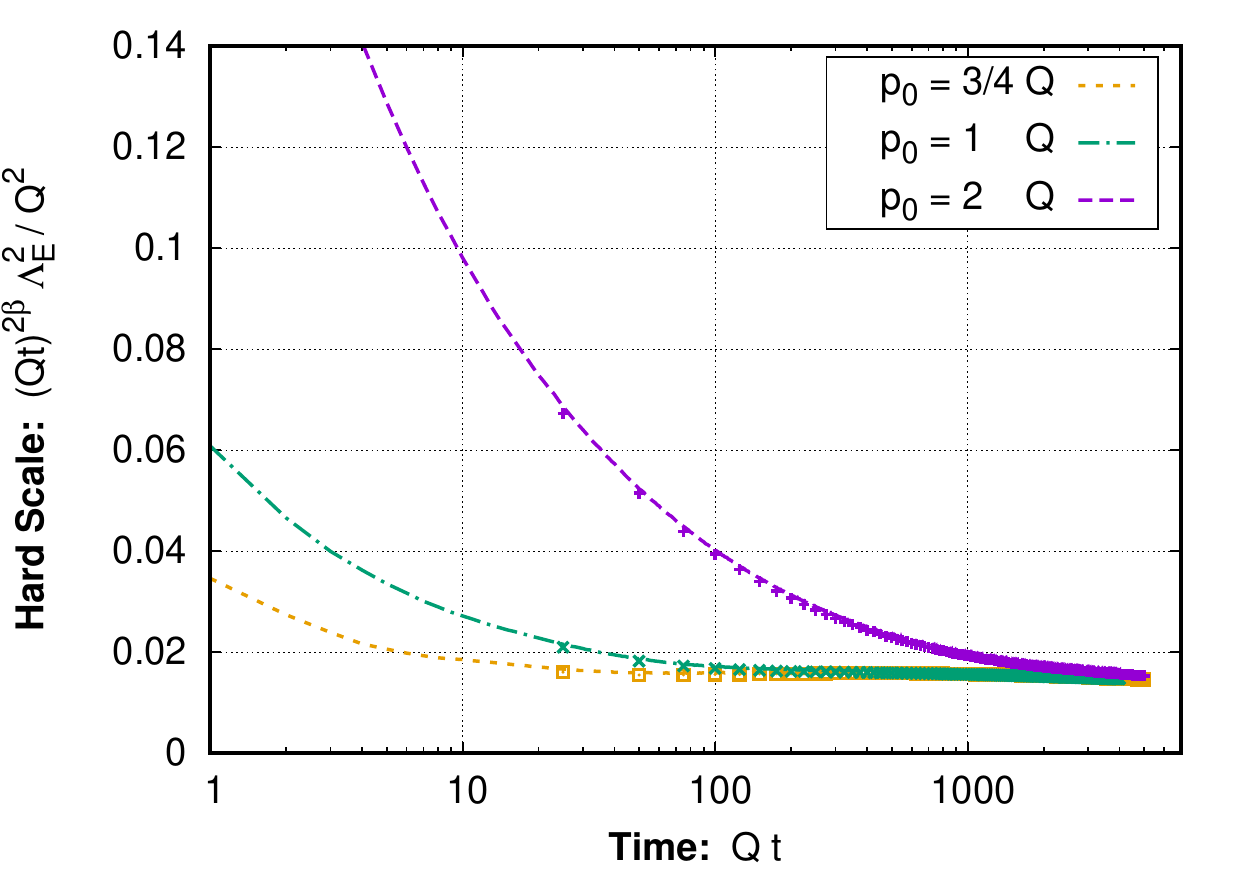}
	\caption{Hard scales $\Lambda^2_E(t)$ for different initial conditions in \tDTwMi\ theory. The  gauge-invariant definition \eqref{eq_hardscale} is shown as dashed lines and the perturbative integral expressions \eqref{eq_hardscale_pert} as points. The curves are rescaled with $t^{2\beta}$ with $\beta = -1/5$.}
	\label{fig_hardscale2}
\end{figure}

\subsection{Gauge-invariant hard scales}
\label{sec:results-gaugeinvobs}

So far, we have observed that the gauge-fixed distribution functions lose their memory on details of their initial conditions and approach a self-similar attractor. To confirm this behavior with gauge-invariant observables, we also compute the time evolution of manifestly gauge-invariant measures of the hard scale \cite{Kurkela:2012hp,Berges:2013eia}
\begin{align}
\label{eq_hardscale}
 \Lambda^2_E(t) \,&= \dfrac{g^2}{\dA \Q^4} \sum_{k,l,i=1,2}\left\langle (D_k^{ab}F^b_{ki}(t,\mbf x))(D_l^{ad}F^d_{li}(t,\mbf x)) \right\rangle \\
 \Lambda^2_\pi(t) \,&= \dfrac{g^2}{\dA \Q^4} \sum_{k,l=1,2}\left\langle (D_k^{ab}D_k^{bc}\phi^c(t,\mbf x))(D_l^{ad}D_l^{de}\phi^e(t,\mbf x)) \right\rangle .
 \label{eq_hardscale_pi}
\end{align}
These provide typical hard momentum scales that are expected to grow as $\Lambda^2(t) \sim t^{-2\beta}$ in the self-similar regime. This can be seen from their perturbative expressions
\begin{align}
\label{eq_hardscale_pert}
 \Lambda^2_{\mrm{pert},E/\pi}(t) = \int \dfrac{\ud^2 p}{(2\pi)^2}\, \dfrac{p^3}{\Q^3}\, \dfrac{g^2 f_{E/\pi}}{\Q} \,,
\end{align}
where all higher powers in the field amplitude  were neglected and the Coulomb gauge condition was used. Note that because of $\Q^4 \propto g^2 \varepsilon$, the hard scales can be interpreted as ratios $\propto \int \ud^2p\, p^2\, \omega(p)f / \int \ud^2p\; \omega(p)f$, characterizing the momentum scale that dominates the  energy density.

The gauge-invariant hard scales $\Lambda^2(t)$, rescaled with $t^{2\beta}$, are shown in \fig\ref{fig_hardscale} in a linear-logarithmic plot as dashed lines for the gauge and scalar sectors of \tDTwMi\ and \tDThrMi\ theories for $\pInit = \Q$ initial conditions. The data points of matching color indicate the respective perturbative expressions $\Lambda^2_{\mrm{pert}}(t)$ that are obtained by integrating the gauge-fixed distribution functions according to \eq\eqref{eq_hardscale_pert}. 
The good agreement between points and lines of the same color confirms our interpretation of the hard scales as $\Lambda^2_{E/\pi} \approx \Lambda^2_{\mrm{pert},E/\pi}$. Moreover, hard scales from different sectors and theories agree well $\Lambda^2_{E} \approx \Lambda^2_{\pi}$. One observes that for $\Q t \gtrsim 75$, the rescaled hard scales become approximately constant, indicating $\Lambda^2_{E/\pi}(t)/\Q^2 \propto (\Q t)^{-2\beta}$. The onset time of self-similar scaling and the value for $\beta$ employed in \fig\ref{fig_hardscale} are the same as used for the self-similar evolution in \fig\ref{fig_selfsim_2D}. A similar power law evolution of the hard scale in \tDThrMi\ theory has been observed in \re\cite{Lappi:2017ckt}. This consistency between gauge-invariant and gauge-fixed observables confirms the emergence of a self-similar attractor. 

In general, the approach to the attractor depends on the initial conditions and on the observables. This is illustrated for the \tDTwMi\ theory in  \fig\ref{fig_hardscale2}, which shows the evolution of the hard scale in the  \tDTwMi\ theory for initial conditions with different values of $\pInit$. 
The flattening of this observable indicates the onset of a power-law evolution. One observes that for the hard scales, the onset of scaling becomes slower with larger $\pInit$ or, equivalently, with lower amplitude $n_0$ (cf., \fig\ref{fig_fp_diffIC_2D}).

\subsection{Longitudinal polarization and Debye mass}
\label{sec:results-corr_mass}

While distribution functions are useful for a comparison to kinetic theory, one can extract further information about a system by studying more general correlation functions. Due to our definition of distribution functions in \eq\eqref{eq_defin_fp}, we have already discussed the evolution and properties of the transversely polarized equal-time correlator $\langle E_T E_T^* \rangle(t,p) \equiv p \fE(t,p)$ and its scalar analogy $\langle \pi \pi^* \rangle(t,p) \equiv p \fpi(t,p)$. Let us now focus on the longitudinally polarized equal-time correlation function $\langle E_L E_L^* \rangle(t,p)$. It is shown in \fig\ref{fig_TvsL} at fixed time $\Q t = 2000$, together with the transverse polarization and the scalar correlator for \tDTwMi\ in the upper and for \tDThrMi\ theory in the lower panel. Every correlation function is shown for two different sets of discretization parameters, written in terms of the lattice length $L = a_s N_s$ and the lattice spacing $a_s$. The good agreement among curves of different volumes and lattice spacings indicates the absence of numerical lattice artefacts. 

The correlators $\langle E_T E_T^* \rangle$ in \fig\ref{fig_TvsL} are flat up to a high momentum $p \sim \Lambda$ beyond which they decrease rapidly, which is, of course, equivalent to our previous observation that $f(t,p) \sim 1/p$ up to a hard scale. Similarly, we have observed the IR enhancement of the $\langle \pi \pi^* \rangle$ correlator and its agreement with $\langle E_T E_T^* \rangle$ at higher momenta already at the example of $\fpi$. The longitudinally polarized correlator $\langle E_L E_L^* \rangle$ approaches $\langle E_T E_T^* \rangle$ at the lowest momenta, while strongly differing for momenta above the Debye mass $m_D$. Indeed, as known in thermal equilibrium \cite{Kurkela:2012hp} and also observed far from equilibrium in the \tDThrIso\ theory \cite{Boguslavski:2018beu}, the longitudinal correlation function follows the form
\begin{align}
\label{eq_fit_long}
 \langle E_L E_L^* \rangle \approx \dfrac{A}{1 + (p^2/m_D^2)^{1 + \delta}}\,,
\end{align}
for momenta $p \lesssim \Lambda$. In the late-time limit and in thermal equilibrium one then expects $\delta = 0$. We have fitted this form to $\langle E_L E_L^* \rangle$ and included the fits in \fig\ref{fig_TvsL} as black dashed lines. They are seen to accurately describe the correlator.

\begin{figure}[t]
	\centering
	\includegraphics[scale=0.7]{\pToFigs/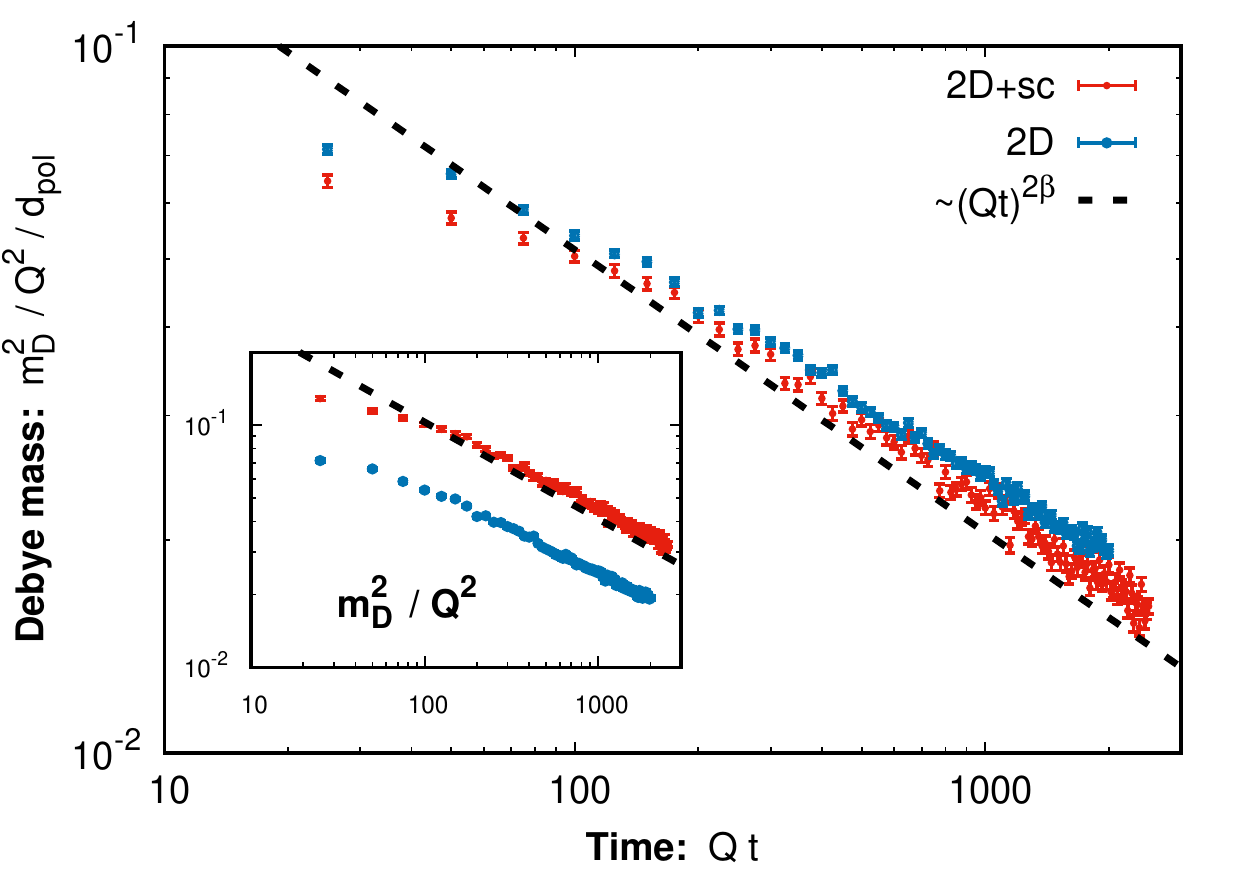}
	\caption{Time evolution of $m_D^2$ extracted from fits of the form \eqref{eq_fit_long} to the longitudinal correlator in \fig\ref{fig_TvsL} and shown for \tDTwMi\ and \tDThrMi\ theories. The black dashed line corresponds to a power law $t^{2\beta}$ with $\beta = -1/5$, leading to \eq\eqref{eq_relation_mD_Lmabda}.}
	\label{fig_mass}
\end{figure}

Fitting this form to the longitudinal correlator at different times, we have extracted the evolution of the fitting parameters $A$, $\delta$ and $m_D$. As expected from our previous discussions, the amplitude quickly approaches an $A \sim t^{\alpha - \beta}$ power law and the deviation $\delta$ monotonously decreases from $\delta \approx 0.2 - 0.3$ at early times to $\delta \approx 0.08 - 0.12$ at time $\Q t = 2000$ for both theories. 

Most interestingly, the fitting procedure allows us to extract an estimate for the Debye mass $m_D$ from the $p$-dependence of the correlator. Its time evolution is shown in \fig\ref{fig_mass}. In the main figure, the normalized $m_D^2/\dpol$ is plotted as a function of time on a double-logarithmic panel. One observes that the curves stemming from the different theories almost coincide while they lie far apart in the inset where $m_D^2$ is depicted. This indicates that $m_D^2$ scales with the number of degrees of freedom $\dpol$, which are $1$ for \tDTwMi\ and $2$ for \tDThrMi\ theory. Moreover, $m_D^2$ is observed to approach a $t^{2\beta}$ power law evolution that is represented by a black dashed line. Its power law evolution sets in roughly at the same time scale as for the hard scale in the upper panel of \fig\ref{fig_hardscale}. These observations suggest a relation 
\begin{align}
\label{eq_relation_mD_Lmabda}
    m_D^2 \;&\sim \dpol\, \dfrac{\Q^4}{\Lambda^2} \nonumber \\
    &\sim \dpol\,g^2 \nha\,\Lambda \,,
\end{align}
where we used energy conservation in the last line and where $\nha$ is the amplitude at hard momenta $p \sim \Lambda$. Since \eq\eqref{eq_relation_mD_Lmabda} leads to $m_D/\Lambda \sim (\Q t)^{2\beta} \ll 1$, the scale separation between the soft scale $m_D$ and the hard scale $\Lambda$ increases with time. This should in general allow for a perturbative (HL) expansion with the ratio $m_D/\Lambda$ as the expansion parameter.

The  leading order HL expression for the Debye mass is \cite{Laine:2016hma}
\begin{align}
 \label{eq_DebyeM}
 m_{D,\HL}^2 &\approx \int \frac{\ud^2 p}{(2\pi)^2} \frac{g^2\left(\nc \fE(t,p) + \nc \fpi(t,p) \right)}{p} \nonumber \\
 \,&= \dpol \nc \int \frac{\ud^2 p}{(2\pi)^2} \frac{g^2f(t,p)}{p}\,,
\end{align}
where $\fpi \equiv 0$ for the \tDTwMi\ theory and where $f(t,p)$ is an average distribution. Because of $f \sim 1/p$ at low momenta, or even steeper for the scalar distribution, the integral diverges in the IR in 2 spatial dimensions  and needs to be regularized by some cutoff at the scale $m_D$. This leads to 
\begin{equation}
m_{D,\HL}^2 \sim g^2\nha\, \Lambda\, \dpol \nc\, \ln ( \Lambda / m_{D,\HL} ),
\end{equation}
 bringing a logarithmic correction to the 
 estimate \eqref{eq_relation_mD_Lmabda}. The HL expression \eqref{eq_DebyeM} was used in \re\cite{Lappi:2017ckt} as one of three methods to extract the mass scale. In all the methods employed, the mass followed an approximate power law evolution with $m_D/\Q \sim (\Q t)^{\beta}$ with values for $\beta$ that are roughly consistent with our results.
We will return to the discussion of the physical interpretation of this logarithm in \se\ref{sec_Boltzmann_class}.


\section{Scaling behavior in a kinetic theory picture}
\label{sec_Boltzmann_class}

The non-equilibrium evolution of gauge theories can also be studied using an effective kinetic theory setup. 
In \cite{Arnold:2002zm}, an effective kinetic theory has been formulated for $d=3$ spatial dimensions which is defined by a Boltzmann transport equation
\begin{align}
 \label{eq_Boltzmann_eq}
 \frac{\partial f(t,p)}{\partial t} = -C[f](t,p)\,,
\end{align}
where $f$ is the distribution function of gluons and where the effective collision kernel $C[f]$ is the sum over the relevant elastic and inelastic scattering processes between the particles. Many of the features of the over-occupied UV-cascading system have been well understood in terms of such a kinetic theory description~\cite{Kurkela:2012hp,Berges:2012ev,York:2014wja}.
This effective kinetic theory describes the evolution of modes at momentum scales well above the screening scale $p \gg m_D$. With the assumption that scattering against modes that carry soft momenta is subdominant compared to the scattering with the hard particles, this effective description may be used to follow the time evolution of the hard scale at late times when a scale separation between the soft and hard scales has developed. 

The soft scale makes its entrance to the kinetic theory because of the Coulomb-divergent $t$- and $u$-channel elastic scattering amplitudes%
\footnote{Here $|M|^2$ is expressed in a non-relativistic normalization related to the usual relativistic normalization by $|M|^2 = |\mathcal{M}|^2/(16\, p\,k\,p'\,k')$.
}
$|M|_{\mrm{vacuum}}^2 \sim g^4/(q_\perp^2 )^2$ appearing in the elastic part of the collision kernel 
\begin{align}
 \label{eq_collInt_2to2}
 &~~C^{2\leftrightarrow 2}[f_{\mbf p}] \nonumber \\
 &=\; \frac{1}{2}  \int_{\mbf k, \mbf p', \mbf k'} \left| M \right|^2 (2\pi)^{d+1} \delta^{d+1}\left(P+K-P'-K'\right) \nonumber \\
 &f_{\mbf p} f_{\mbf k} \left(1 + f_{\mbf p'}\right) \left(1 + f_{\mbf k'}\right) - \left(1 + f_{\mbf p}\right) \left(1 + f_{\mbf k}\right) f_{\mbf p'} f_{\mbf k'},
\end{align}
with $\int_{\mbf k} \equiv \int \ud^d k/(2\pi)^d$, $f_{\mbf p} \equiv f(t,\mbf p)$ and ($d+1$)-momenta $P$.
In medium, the Coulomb divergences are regulated by the physics of screening, taking place at the momentum transfer scale $q_\perp \sim m_D$. In $d=3$ dimensions, the particles at the hard scale screen the most, and the scale separation between the soft and the hard scales allows one to describe the screening in simple effective theory, the Hard-Loop effective theory. Because of this simplification, the effective screened matrix elements may be solved analytically
to complete the effective kinetic theory. 

Similarly, in $d=2$ dimensions,  the soft and hard scales are parametrically separated at late times allowing for a quasiparticle description of the hard modes. In contrast to three dimensions, however, the equation for the Debye mass \eqref{eq_DebyeM} in two dimensions gets equal contributions from all momentum scales $m_D < p < \Lambda$, such that soft modes contribute equally to screening. This implies that the modes at the soft scale $m_D$ interact among each other in a non-perturbative way, reminiscent of the ultra-soft \emph{magnetic} scale in three spatial dimensions. The practical implication of this is that the dynamics of the soft modes is no longer governed by a simple effective Hard-Loop theory. Because of this complication, it is not straightforward to analytically find the forms of the required effective matrix elements and formulate a leading-order accurate kinetic theory.

Nevertheless, while we do not have access to a formulation of the kinetic theory that would be accurate at a numerical level, we may still include parametric considerations in this picture following \re\cite{Kurkela:2011ti}. In particular, even while we do not know analytically how the effective elastic scattering matrix element looks like, the elastic scattering amplitude for $t$-channel is regulated by the soft scale and to parametric accuracy is
\begin{align}
 |M|^2 \sim \frac{g^4}{(q_\perp^2 + m_D^2 )^2},
\end{align}
where $q_\perp$ is the transverse momentum transfer and the regulator $m_D$ is at the correct scale but its form is simplified and not correct at a numerical level. 
The hard particles moving in the medium experience then soft scatterings with other particles in the medium with a (soft) elastic scattering rate of
\begin{align}
\label{eq:dgamma}
    \frac{\ud\Gamma}{\ud q_\perp} \sim \frac{ g^4 }{(q_\perp^2 + m_D^2 )^2} \int \ud^2 p \, f (1+f)\,.
\end{align}
This is a 2-dimensional version of the usual kinetic theory relation expressing the rate in terms of the cross section and number density of scattering targets. In two dimensions there is only one transverse direction, and thus the squared amplitude gives a rate differential in a 1-dimensional transverse momentum $\ud q_\perp$. The factor $\int \ud^2p f$ accounts for the number density particles in the medium against whom the collision occurs and $(1+f) \sim f$ is the final state Bose enhancement factor. For a thermal-like infrared spectrum $f\sim T_*/p$ with an effective temperature $T_*$, the integral over $\ud p$ gets equal contributions from all scales, such that collisions with soft particles are equally frequent as those with hard ones. This is in contrast to \tDThrIso, where because of the higher dimensionality and larger phase space at high $p$ most of the scatterings take place against hard particles. Because the kinetic theory describes the evolution of hard modes only, the kinetic theory framework does not numerically describe the collisions against the soft particles. We will, however, neglect this complication here and proceed with our analysis assuming that $p\sim \Lambda$ in  \eq\nr{eq:dgamma},  with the understanding that our accuracy is purely parametric. 

The repeated collisions with the medium particles lead to an integrated momentum transfer of $\Delta p^2 \sim \hat q t$, with the momentum diffusion coefficient of 
\begin{align}
\label{eq_qhat_int}
    \hat q &\sim \int \ud q_\perp \frac{\ud\Gamma}{\ud q_\perp}  q_\perp^2
    \\
    & \sim g^4 \int \ud q_\perp  \frac{  \, q_\perp^2 }{(q_\perp^2 + m_D^2 )^2} \int \ud^2 p f(1+ f).    
\end{align}
In two dimensions  the integral over the momentum transfer $q_\perp$ is dominated by the softest collisions with $q_\perp \sim m_D$. Using further that $m_D^2 \sim g^2 \nha\,\Lambda$ from \eq\eqref{eq_relation_mD_Lmabda}, and reminding the reader that the coupling is dimensionful in \tDTwMi\ with $[g^2] = 1$, the momentum broadening coefficient is parametrically of order
\begin{align}
\label{eq_qhat_param}
    \hat q \sim \frac{\Lambda^2\, (g^2\nha)^2}{m_D} \sim \Lambda^{3/2}\, (g^2 \nha)^{3/2}.
\end{align}
Energy conservation dictates a relation between the amplitude and the hard scale
\begin{align}
\label{eq_nha_param}
    g^2\nha \sim \dfrac{\Q^4}{\Lambda^3}\,,
\end{align}
corresponding to $\alpha = 3 \beta$. With this, the momentum transport coefficient at a given value of the hard scale reads
\begin{align}
\label{qhat}
    \hat q \sim \dfrac{\Q^6}{\Lambda^3}\,.
\end{align}
During the non-equilibrium cascade, elastic scatterings push hard scales to harder momenta and the highest momenta reached at time $t$ are given by
\begin{align}
\label{eq_diff_equ}
    \Lambda^2 \sim \hat q \, t.
\end{align}
This equation governs the evolution of the hard scale. Combining it with \eq\eqref{qhat} gives the time evolution of the hard scale
\begin{align}
    \Lambda \sim \Q\; (\Q t)^{1/5},
\end{align}
corresponding to the value $\beta = -1/5$. These values for the scaling exponents are summarized in \eq\eqref{eq_scaling_values_plots} and were used in all the plots of \se\ref{sec:results}. They are consistent with the numerically determined values in \eq\eqref{eq_scaling_data2}. 

We expect that, as in \tDThrIso\ \cite{Kurkela:2011ti}, the inelastic scattering rate at the hard scale plays an important role as well. 
An important qualitative feature of inelastic scatterings is that they do not conserve particle number. As a consequence, there is no strong build-up of particle number in the infrared.

While the UV-cascade shares commonalities with the one in \tDThrIso, we emphasize one important difference between \tDTwMi\ and \tDThrIso. 
In \tDThrIso, elastic scatterings of large ($q_\perp \sim \Lambda$) and of small ($q_\perp \sim m_D$) momentum transfers both lead to the same scaling exponent $\beta = -1/7$. This would also be the case in \tDTwMi\ if hard scatterings with $q_\perp \sim \Lambda$ were dominant. 
This can be seen from \eq\eqref{eq_qhat_int}, where the hard scales  $q_\perp  \sim \Lambda$  give a contribution $\sim \Lambda\, (g^2\nha)^2 $ to the estimate \eqref{eq_qhat_param}. If this contribution from the hard scales were the dominant one for $\hat q$,  the resulting $\hat q \sim \Lambda\, (g^2\nha)^2 $  together with \eqs\eqref{eq_nha_param} and \eqref{eq_diff_equ} would lead to  $\Lambda \sim \Q\; (\Q t)^{1/7}$ as in \tDThrIso. Such a hard scattering dominance is  clearly disfavored both  by our numerical results and the above analytical arguments.  Instead, collisions with small momentum transfer dominate the momentum diffusion, resulting in a different value for $\beta$.

Since the \tDThrMi\ theory results from a dimensionally reduced three-dimensional theory, and the scalar field is identified with the originally third component of the gauge field $\phi \equiv A_3$, we expect similar arguments to apply for the scalar distribution at high momenta.


\section{Conclusion}
\label{sec:conclu}

We have studied a far-from-equilibrium attractor in 2-dimensional highly occupied gauge theories using real-time classical field simulations. We found that these systems exhibit self-similar evolution of the distribution function that corresponds to an energy cascade towards higher momenta and that its scaling properties can be understood using parametric estimates in kinetic theory.

In particular, we have studied the time-dependence of equal-time field correlators and that of the hard scale $\Lambda$ and of the screening scale $m_D$. We have used different observables, including manifestly gauge-invariant measures, to extract the scaling exponents. We found that both the \tDTwMi\ and \tDThrMi\ theories exhibit self-similar cascades that bring energy towards the UV and are insensitive to the initial conditions. The cascades are characterized by the evolution of the hard scale $\Lambda\sim t^{-\beta}$, whose time evolution is described by a scaling exponent $\beta=-1/5$. Moreover, the Debye scale is extracted from a longitudinally polarized correlator of chromo-electric fields and is shown to decrease with time as $m_D \sim t^{\beta}$ towards low momenta. 
While in \tDTwMi, we do not have access to a leading-order accurate kinetic theory description, these scaling exponents can be understood in terms of parametric consideration within a kinetic theory setup. A crucial difference to three dimensions is that soft scattering is enhanced compared to the hard scattering. Therefore unlike in \tDThrIso, in order to derive the correct scaling exponents, screening effects have to be taken into account. 

These findings are consistent with a description of hard momentum modes in terms of quasi-particle degrees of freedom even if a Hard-Loop theory is insufficient to describe the dynamics of soft modes $\sim m_D$. To learn more about soft dynamics, further numerical studies are required. These include also unequal-time correlators
that can be studied numerically with methods that have been developed recently~\cite{Kurkela:2016mhu} and successfully used for 3-dimensional systems \cite{Boguslavski:2018beu,PineiroOrioli:2018hst}. We will report on results from such studies in a subsequent work. In addition, it would be interesting to better understand the origin of the observed IR enhanced region of the scalar field correlator.

With regard to heavy-ion collision phenomenology, this attractor might emerge at times that are too late to be reached in a collision, since other phenomena like plasma instabilities can set in earlier \cite{Mrowczynski:1994xv,Arnold:2003rq,Rebhan:2004ur,Romatschke:2005pm,Arnold:2007tr,Berges:2012iw,Berges:2012cj}. However, our observation that the evolution of the self-similar attractor can be understood in terms of kinetic estimates is relevant nonetheless for the understanding of the dynamics at early times in heavy-ion collisions: a kinetic description, and thus a description in terms of quasi-particles, can be used to describe two-dimensional plasmas despite the break-down of hard loop resummations. Hence, we show that kinetic descriptions can be valid already at the early times of the evolution of the two-dimensional Glasma. The exact time when such descriptions become valid can be estimated in further studies.


\begin{acknowledgments}
  We are grateful to P.\ Arnold, J.\ Berges, A.\ Mazeliauskas, A.\ Rebhan, P.\ Romatschke, S.\ Schlichting, M.\ Strickland and R.\ Venugopalan for valuable discussions. This project has received funding from the European Research Council (ERC) under the European Union’s Horizon 2020 research and innovation programme (grant agreement No ERC-2015-CoG-681707). The content of this article does not reflect the official opinion of the European Union and responsibility for the information and views expressed therein lies entirely with the authors. 
  K.\ B.\ and J.\ P.\ would like to thank the CERN Theory group for hospitality during part of this work.
  The authors wish to acknowledge CSC - IT Center for Science, Finland, for computational resources. 
\end{acknowledgments}


\appendix

\section{Note on initial conditions in Coulomb gauge}
\label{app_IC_in_CoulombGauge}

In \re\cite{Lappi:2017ckt}, large values for the amplitude $n_0 \gtrsim 1$ were used and it was observed that the initial state changes considerably after the gauge fixing procedure was used. This problem is primarily caused by the nonlinear mapping between $\su(\nc)$ algebra elements $A_k$ and $\SU(\nc)$ group elements $U_k$. In this work, we circumvent this problem in several ways simultaneously. First of all, we employ smaller initial amplitudes $n_0 < 1$. Secondly, we construct the initial link field $U_k(\mbf x)$ in such a way that its Fourier transformed anti-Hermitian traceless part $[U_k]_\ah(\mbf p)$ \footnote{This is defined by $\left[V\right]_\ah \equiv \frac{-i}{2} \left( V- V^\dagger - \frac{1}{\nc} \Tr\left(V - V^\dagger\right)\right)$ for an $\SU(\nc)$ matrix $V$.} 
(and not the logarithm as in~\re\cite{Lappi:2017ckt}) of the link is constructed to reproduce the desired momentum distribution~\nr{eq_2D_IC}. Thus here
\begin{align}
 -g^{jk} \partial_j [U_k]_\ah(\mbf x) = 0
\end{align}
is correct to machine precision initially without the need of an additional gauge fixing procedure. This avoids the issue of the exponentiation of the algebra element spoiling the transversality of the field that was problematic in the study~\cite{Lappi:2017ckt}.

Also note that since no fields depend on the $x^3$ coordinate, the Coulomb gauge condition only applies to $k = 1, 2$, while the adjoint scalars of theory \tDThrMi\ do not enter the condition.

\section{Scaling exponents from likelihood analysis}
\label{app:likelihood}

To extract the scaling exponents $\alpha$ and $\beta$, we employ the self-similarity analysis of \re\cite{Berges:2013fga} and its modification \cite{Orioli:2015dxa} for the self-similar evolution of the \tDTwMi\ theory displayed in \fig\ref{fig_selfsim_2D}. We define a rescaled distribution
\begin{align}
    f_{\mrm{test}}(t,p) = (t/t_r)^{-\alpha} f\left(t, (t/t_r)^{-\beta} p\right).
\end{align}
By construction, this rescaled distribution is $f_{\mrm{test}}(t_r,p) \equiv f(t_r,p)$ for the reference time $\Q t_r = 500$. In case of self-similarity, $f_{\mrm{test}}(t,p)$ is time-independent. Hence, its difference to the distribution at $t_r$ is a good measure of deviation from a self-similar evolution. We can quantify the deviation by computing
\begin{align}
    \chi^2_{m}(\tilde{\alpha},\beta) = \dfrac{1}{N_t} \sum_{i} \dfrac{\int \ud \log p\,\left(p^m \Delta f(t_i,p)\right)^2}{\int \ud \log p\,(p^m f(t_r,p))^2},
\end{align}
with $\Delta f(t_i,p) = f_{\mrm{test}}(t_i,p) - f(t_r,p)$ and with the exponent of the energy density $\tilde{\alpha} \equiv \alpha - 3\beta$. Momentum integrals are performed in the interval $0.2 \leq p/\Q \leq 5$. The deviations $\chi^2_{m}$ are averaged over the test times $\Q t_i = 75, 200, 1500, 4000, 16000$ used in \fig\ref{fig_selfsim_2D} for different moments with $m = 2, \dots, 5$. For brevity, we will omit the index $m$. We can now define a likelihood function
\begin{align}
    W(\tilde{\alpha},\beta) = \dfrac{1}{\mathcal{N}}\, \exp\left(1 - \dfrac{\chi^2(\tilde{\alpha},\beta)}{\chi^2_{\mrm{min}}} \right),
\end{align}
where $\chi^2(\tilde{\alpha}_0,\beta_0) \equiv \chi^2_{\mrm{min}}$ takes its minimal value. The normalization $\mathcal{N}$ is chosen to satisfy $\int \ud \tilde{\alpha}\, \ud \beta\; W(\tilde{\alpha},\beta) = 1$. We integrate over one of the exponents to obtain an estimate for the distribution of the other exponent, e.g., $W(\beta) = \int \ud \tilde{\alpha} \, W(\tilde{\alpha},\beta)$. We extract an estimate for the uncertainty $\sigma_\beta$ for every $m$ by fitting the resulting distributions to Gaussian functions $\propto \exp[- (\beta-\beta_0)^2 / (2 \sigma_\beta^2)]$. 

The statistical error $\sigma_{\beta}^{\chi}$ of the $\chi^2$ fit is estimated as the maximal value of $\sigma_\beta$ among the different $m$, giving $\sigma_{\beta}^{\chi} = 0.012, \ \sigma_{\tilde{\alpha}}^{\chi} = 0.019$. We can also extract a systematical error by varying  $m$  and requiring that all $\beta_0$ values for different values of $m$ lie in the interval $[\bar{\beta}_0 - \sigma_{\beta}^{\mrm{sys}}, \bar{\beta}_0 + \sigma_{\beta}^{\mrm{sys}}]$ and similarly for $\tilde{\alpha}$. This leads to the error estimates $\sigma_{\beta}^{\mrm{sys}} = 0.004$ and $\sigma_{\tilde{\alpha}}^{\mrm{sys}} = 0.0035$.
The statistical  $\chi^2$ errors are clearly the larger of these. The mean values and error estimates quoted in \eqs\eqref{eq_scaling_data}, \eqref{eq_scaling_data2} are obtained by combining and rounding the mean values and error estimates.


\bibliographystyle{JHEP-2modlong}
\bibliography{spires}

\end{document}